\documentclass[prd,showpacs,nofootinbib,showkeys]{revtex4}

\usepackage{graphicx}
\usepackage{amsmath}
\usepackage{amssymb}
\usepackage{amsxtra}

\begin{document}

\title{Charged lepton and Neutrino masses from a low energy $SU(3)$ flavour symmetry model}

\author{Albino Hern\'andez-Galeana}

\email{albino@esfm.ipn.mx}

\affiliation{ Departamento de F\'{\i}sica,   Escuela Superior de
F\'{\i}sica y Matem\'aticas, I.P.N., \\
U. P. "Adolfo L\'opez Mateos". C. P. 07738, M\'exico, D.F., M\'exico. }



\begin{abstract}
We report a recent study on lepton masses
within a beyond standard model with a $SU(3)$ family symmetry model.
In this scenario ordinary heavy fermions, top and bottom quarks and tau
lepton become massive at tree level from {\bf Dirac See-saw} mechanisms
implemented by the introduction of a new set of $SU(2)_L$ weak singlet vector-like
fermions $U,D,E,N$, with $N$ a sterile neutrino. The $N_{L,R}$ sterile neutrinos allow
the implementation of a $8\times 8$ general tree level See-saw Majorana neutrino mass
matrix with four massless eigenvalues. Hence, light fermions, including
light neutrinos, obtain masses from one loop radiative corrections mediated
by the massive $SU(3)$ gauge bosons. Recent analysis shows the existence of a
space parameter region where the $M_U, M_D,M_E,M_N$ vector-like fermion masses
are within a scale of a few TeV's. This BSM model is able to accommodate the
known spectrum of quark masses and mixing in a $4\times 4$
non-unitary $V_{CKM}$ as well as the charged lepton masses.
We report a preliminary solution for ordinary charged lepton masses
$( m_e \,,\, m_\mu \,,\, m_\tau ) = ( 0.486 \,,\,102.7\,,\,1746.17 )\,\text{MeV}$
at the $M_Z$ scale, a $M_E \approx 2.6 \,\text{TeV}$ and $SU(3)$ gauge boson masses
of $\mathcal{O} (10\,\text{TeV})$.

A detailed numerical analysis on neutrino masses and mixing is in progress
and hopefully shall be reported soon.
\end{abstract}

\maketitle

\tableofcontents

\section{ Introduction }

Recently several experiments have reported new experimental results on
neutrino mixing\cite{Altarelli.Smirnov}, on large $\theta_{13}$ mixing from Daya Bay\cite{DayaBay},
T2K\cite{T2K}, MINOS\cite{MINOS}, DOUBLE CHOOZ\cite{DOUBLE}, and RENO\cite{RENO}, implying
a deviation from TBM\cite{TBM} scenario, including possible evidence for the existence of
sterile neutrinos from LSND and MiniBooNE\cite{MiniBooNE,LSND-MiniBooNe}.

The strong hierarchy of quark and charged lepton masses and quark mixing
have suggested to many model building
theorists that light fermion masses could be generated from
radiative corrections\cite{earlyradm}, while those of the top and
bottom quarks as well as that of the tau lepton are generated at
tree level. This may be understood as a consequence of the
breaking of a symmetry among families ( a horizontal symmetry ).
This symmetry may be discrete \cite{modeldiscrete}, or continuous,
\cite{modelcontinuous}. The radiative generation of the light
fermions may be mediated by scalar particles as it is proposed,
for instance, in references \cite{modelrad,medscalars} and the
author in \cite{prd2007}, or also through vectorial bosons as it
happens for instance in "Dynamical Symmetry Breaking" (DSB) and
theories like " Extended Technicolor " \cite{DSB}.

\vspace{2mm}

In this report we address the problem of fermion masses and
quark mixing within an extension of the SM introduced by the
author in \cite{albinosu32004}, which includes a $SU(3)$\cite{su3models} gauged flavor
symmetry commuting with the SM group. In previous
reports\cite{albinosu32009} we showed that this model has the
properties to accommodate a realistic spectrum of charged fermion
masses and quark mixing. We introduce a hierarchical mass
generation mechanism in which the light fermions obtain masses
through one loop radiative corrections, mediated by the massive
bosons associated to the $SU(3)$ family symmetry that is
spontaneously broken, while the masses for the top and bottom
quarks as well as for the tau lepton, are generated at tree level
from "Dirac See-saw"\cite{SU3MKhlopov} mechanisms implemented by
the introduction of a new generation of $SU(2)_L$ weak singlets
vector-like fermions. Recently, some authors have pointed out
interesting features regarding the possibility of the existence of
a sequential fourth generation\cite{fourthge}. Theories and models with extra
matter may also provide interesting scenarios for present cosmological problems,
such as candidates for the nature of the Dark Matter
(\cite{normaapproach},\cite{khlopov}). This is the case of an extra
generation of vector-like matter, both from theory and current
experiments\cite{vector-like-SU(2)-weak-singlets}. Due to the fact that the
vector-like quarks do not couple to the $W$ boson, the mixing of $U$ and $D$
vector-like quarks with the SM quarks yield an extended $4\times
4$ non-unitary CKM quark mixing matrix. It has pointed out for
some authors that these type of vector-like fermions are weakly constrained from
Electroweak Precision Data (EWPD) because they do not break
directly the custodial symmetry, then main experimental
constraints on vector-like matter come from the direct production
bounds and their implications on flavor physics. See ref.
\cite{vector-like-SU(2)-weak-singlets} for further details on
constraints for $SU(2)_L$ singlet vector-like fermions.

\vspace{3mm}

\section{Model with $SU(3)$ flavor symmetry}

\subsection{Fermion content}

We define the gauge group symmetry $G\equiv SU(3) \otimes G_{SM}$
, where $SU(3)$ is a flavor symmetry among families and
$G_{SM}\equiv SU(3)_C \otimes SU(2)_L \otimes U(1)_Y$ is the
"Standard Model" gauge group, with $g_s$, $g$ and $g^\prime$ the corresponding
coupling constants. The content of fermions assumes the ordinary quarks and
leptons assigned under G as:

\begin{equation*}
\psi_q^o = ( 3 , 3 , 2 , \frac{1}{3} )_L  \qquad ,\qquad \psi_u^o = ( 3 , 3, 1 , \frac{4}{3} )_R
\qquad ,\qquad \psi_d^o = (3, 3 , 1 , -\frac{2}{3} )_R  \end{equation*}

\begin{equation*} \psi_l^o =( 3 , 1 , 2 , -1 )_L \qquad , \qquad \psi_e^o = (3, 1 , 1,-2)_R  \;, \end{equation*}

\vspace{3mm}
\noindent where the last entry corresponds to the
hypercharge $Y$, and the electric charge is defined by $Q = T_{3L}
+ \frac{1}{2} Y$. The model also includes two types of extra
fermions: Right handed neutrinos $\Psi_\nu^o = ( 3 , 1 , 1 , 0
)_R$, and the $SU(2)_L$ singlet vector-like fermions

\begin{equation}
U_{L,R}^o= ( 1 , 3 , 1 , \frac{4}{3} )  \qquad , \qquad D_{L,R}^o
= ( 1 , 3 , 1 ,- \frac{2}{3} )  \label{vectorquarks}
\end{equation}

\begin{equation}
N_{L,R}^o= ( 1 , 1 , 1 , 0 )
\qquad , \qquad E_{L,R}^o= ( 1 , 1 , 1 , -2 ) \label{vectorleptons} \end{equation}

The above fermion content and its assignment under the group G
make the model anomaly free. After the definition of the gauge
symmetry group and the assignment of the ordinary fermions in the
canonical form under the standard model group and in the
fundamental $3$-representation under the $SU(3)$ family symmetry,
the introduction of the right-handed neutrinos is required to
cancel anomalies\cite{T.Yanagida1979}. The $SU(2)_L$ weak singlets
vector-like fermions have been introduced to give masses at tree
level only to the third family of known fermions through Dirac
See-saw mechanisms. These vector like fermions play a crucial role
to implement a hierarchical spectrum for quarks and charged lepton
masses together with the radiative corrections.

\section{Electroweak symmetry breaking}

Recently ATLAS\cite{ATLAS} and CMS\cite{CMS} at the Large Hadron Collider announced
the discovery of a Higgs-like particle, whose properties, couplings to fermions
and gauge bosons will determine whether it is the SM Higgs or a member of an extended
Higgs sector associated to an BSM theory.  The electroweak symmetry breaking in the
$SU(3)$ family symmetry model involve the introduction of two triplets of $SU(2)_L$
Higgs doublets.

\vspace{4mm}
To achieve the spontaneous breaking of the electroweak symmetry to
$U(1)_Q$,  we introduce the scalars: $\Phi^u = ( 3 , 1 , 2 , -1 )$
and $\Phi^d = ( 3 , 1 , 2 , +1 )$, with the VEV´s: $\langle
\Phi^u \rangle^T = ( \langle \Phi_1^u \rangle , \langle \Phi_2^u \rangle
, \langle \Phi_3^u \rangle )$ , $\langle \Phi^d \rangle^T = (
\langle \Phi_1^d \rangle , \langle \Phi_2^d \rangle
, \langle \Phi_3^d \rangle )$, where $T$ means transpose,
and

\begin{equation} \qquad \langle \Phi_i^u \rangle = \frac{1}{\sqrt[]{2}} \left(
\begin{array}{c} v_i
\\ 0  \end{array} \right) \qquad , \qquad
\langle \Phi_i^d \rangle = \frac{1}{\sqrt[]{2}} \left(
\begin{array}{c} 0
\\ V_i  \end{array} \right) \:.\end{equation}

\vspace{3mm}
\noindent The contributions from
$\langle \Phi^u \rangle$ and $\langle \Phi^d \rangle$ yield
to the $W$ and $Z$ gauge boson masses

\begin{equation} \frac{g^2 }{4} \,(v_u^2+v_d^2)\,
W^{+} W^{-} + \frac{ (g^2 + {g^\prime}^2) }{8}  \,(v_u^2+v_d^2)\,Z_o^2   \end{equation}

\noindent $v_u^2=v_1^2+v_2^2+v_3^2$ , $v_d^2=V_1^2+V_2^2+V_3^2 $.  Hence,
if we define as usual $M_W=\frac{1}{2} g v$, we may write $ v=\sqrt{v_u^2+v_d^2 } \thickapprox
246$ GeV.

\section{$SU(3)$ flavor symmetry breaking}

To implement a hierarchical spectrum for charged fermion masses,
and simultaneously to achieve the SSB of $SU(3)$, we introduce the
flavon scalar fields: $\eta_i,\;i=2,3$, transforming under the gauge
group as $(3 , 1 , 1 , 0)$ and taking the "Vacuum Expectation
Values" (VEV's):

\begin{equation}
\langle \eta_2 \rangle^T = ( 0 , \Lambda_2,0)  \quad , \quad
\langle \eta_3 \rangle^T = ( 0 , 0, \Lambda_3)
\end{equation}

\noindent The above scalar fields and VEV's break completely the
$SU(3)$ flavor symmetry. The corresponding $SU(3)$ gauge bosons
are defined in Eq.(\ref{SU3lagrangian}) through their couplings to
fermions. Thus, a natural hierarchy among the VEV´s scales  is
$ \Lambda_2\:,\:\Lambda_3 \; \gg v \simeq 246\:\text{GeV}$. Therefore,
neglecting tiny contributions from electroweak symmetry breaking,
we obtain in good approximation the gauge bosons mass terms

\begin{eqnarray} \frac{M_1^2}{2} \,Z_1^2 + (\frac{4}{3} \,M_2^2 + \frac{1}{3} \,M_1^2) \frac{Z_2^2}{2}
- \frac{M_1^2}{ \sqrt{3} } \,Z_1 \,Z_2     \nonumber \\
                                           \nonumber \\
 M_1^2 \,Y_1^+ Y_1^- + M_2^2 \,Y_2^+ Y_2^- + ( M_1^2 +  M_2^2 ) \,Y_3^+ Y_3^-
\end{eqnarray}

\begin{equation} M_1^2=\frac{g_{H_2}^2 \Lambda_2^2}{2} \qquad , \qquad  M_2^2= \frac{g_{H_3}^2 \Lambda_3^2}{2} \qquad , \qquad M_3^2=M_1^2 + M_2^2
\label{M1M2} \end{equation}

\noindent The diagonalization of the $Z_1-Z_2$ squared mass matrix yields the eigenvalues

\begin{eqnarray}
M_-^2=\frac{2}{3} \left( M_1^2 + M_2^2 - \sqrt{ (M_2^2 -  M_1^2)^2+ M_1^2 M_2^2 } \right) \nonumber \\
                                                                                  \nonumber \\
M_+^2=\frac{2}{3} \left( M_1^2 + M_2^2 + \sqrt{ (M_2^2 -  M_1^2)^2+ M_1^2 M_2^2 } \right)
\label{MmMp} \end{eqnarray}

\noindent and therefore the gauge boson masses

\begin{equation}  M_-^2 \,\frac{Z_-^2}{2} +  M_+^2 \,\frac{Z_+^2}{2} +  M_1^2 \,Y_1^+ Y_1^- + M_2^2 \,Y_2^+ Y_2^- + ( M_1^2 +  M_2^2 ) \,Y_3^+ Y_3^-
\end{equation}

\noindent where

\begin{equation}
\begin{pmatrix} Z_1 \\ Z_2  \end{pmatrix} = \begin{pmatrix} \cos\phi & - \sin\phi \\
\sin\phi & \cos\phi  \end{pmatrix} \begin{pmatrix} Z_- \\ Z_+  \end{pmatrix}
\end{equation}

\begin{equation*} \cos\phi \, \sin\phi=\frac{\sqrt{3}}{4} \, \frac{ M_1^2}{\sqrt{ (M_2^2 -  M_1^2)^2+ M_1^2 M_2^2 }}
\end{equation*}

\noindent with the hierarchy $M_1 , M_2 , M_- ,M_+ \gg M_W$.

\section{ Fermion masses}

\subsection{Dirac See-saw mechanisms}

Now we describe briefly the procedure to get the masses for
fermions. The analysis is presented explicitly for the charged
lepton sector, with a completely analogous procedure for the u
and d quarks. With the fields of particles introduced in
the model, we may write the gauge invariant Yukawa couplings

\begin{equation}
h\:\bar{\psi}_l^o \:\Phi^d \:E_R^o \;+\;
h_2 \:\bar{\psi}_e^o \:\eta_2 \:E_L^o \;+\; h_3 \:\bar{\psi}_e^o
\:\eta_3 \:E_L^o \;+\; M \:\bar{E}_L^o \:E_R^o \;+
h.c \label{DiracYC} \end{equation}

\noindent where $M$ is a free mass parameter ( because its mass
term is gauge invariant) and $h$, $h_1$, $h_2$ and $h_3$ are
Yukawa coupling constants.

When the involved scalar fields acquire
VEV's we get, in the gauge basis ${\psi^{o}_{L,R}}^T = ( e^{o} ,
\mu^{o} , \tau^{o}, E^o )_{L,R}$, the mass terms $\bar{\psi}^{o}_L
{\cal{M}}^o \psi^{o}_R + h.c $, where

\begin{equation}
{\cal M}^o = \begin{pmatrix}
0 & 0 & 0 & h \:v_1\\
0 & 0 & 0 & h \:v_2\\
0 & 0 & 0 & h \:v_3\\
0 & h_2 \Lambda_2 & h_3 \Lambda_3 & M \end{pmatrix} \equiv
\begin{pmatrix}
0 & 0 & 0 & a_1\\
0 & 0 & 0 & a_2\\
0 & 0 & 0 & a_3\\
0 & b_2 & b_3 & c \end{pmatrix}
 \; \label{tlmassmatrix} \end{equation}

\noindent Notice that ${\cal{M}}^o$ has the same structure of a
See-saw mass matrix, here for Dirac fermion masses.
So, we call ${\cal{M}}^o$ a {\bf "Dirac See-saw"} mass matrix.
${\cal{M}}^o$ is diagonalized by applying a biunitary
transformation $\psi^{o}_{L,R} = V^{o}_{L,R} \;\chi_{L,R}$. The
orthogonal matrices $V^{o}_L$ and $V^{o}_R$ are obtained
explicitly in the Appendix A. From $V_L^o$ and $V_R^o$, and using
the relationships defined in this Appendix, one computes

\begin{eqnarray}
{V^{o}_L}^T {\cal{M}}^{o} \;V^{o}_R =Diag(0,0,-
\sqrt{\lambda_-},\sqrt{\lambda_+})   \label{tleigenvalues}\\
                                  \nonumber   \\
{V^{o}_L}^T {\cal{M}}^{o} {{\cal{M}}^{o}}^T \;V^{o}_L = {V^{o}_R}^T
{{\cal{M}}^{o}}^T {\cal{M}}^{o} \;V^{o}_R =
Diag(0,0,\lambda_-,\lambda_+)  \:.\label{tlLReigenvalues}\end{eqnarray}

\noindent where $\lambda_-$ and $\lambda_+$ are the nonzero
eigenvalues defined in Eqs.(\ref{nonzerotleigenvalues}-\ref{paramtleigenvalues}),
$\sqrt{\lambda_+}$ being the fourth heavy fermion mass, and $\sqrt{\lambda_-}$ of
the order of the top, bottom and tau mass for u, d and e fermions, respectively.
We see from Eqs.(\ref{tleigenvalues},\ref{tlLReigenvalues}) that at tree level the
See-saw mechanism yields two massless eigenvalues associated to the light fermions:

\subsection{One loop contribution to fermion masses}

Subsequently, the masses for the light fermions arise through one
loop radiative corrections. After the breakdown of the electroweak
symmetry we can construct the generic one loop mass diagram of
Fig. 1. Internal fermion line in this diagram represent the Dirac see-saw
mechanism implemented by the couplings in Eq.(\ref{DiracYC}). The vertices
read from the $SU(3)$ flavor symmetry interaction Lagrangian

\begin{multline} i {\cal{L}}_{int} = \frac{g_{H}}{2}
\left( \bar{e^{o}}
\gamma_{\mu} e^{o}- \bar{\mu^{o}} \gamma_{\mu} \mu^{o} \right) Z_1^\mu
+  \frac{g_{H}}{2 \sqrt{3}} \left( \bar{e^{o}} \gamma_{\mu} e^{o}+ \bar{\mu^{o}}
\gamma_{\mu} \mu^{o} - 2 \bar{\tau^{o}}
\gamma_{\mu} \tau^{o}  \right) Z_2^\mu                \\
+ \frac{g_{H}}{\sqrt{2}} \left( \bar{e^{o}} \gamma_{\mu} \mu^{o} Y_1^{+} +
\bar{e^{o}} \gamma_{\mu} \tau^{o} Y_2^{+} + \bar{\mu^{o}} \gamma_{\mu} \tau^{o} Y_3^{+} + h.c.
\right) \:,\label{SU3lagrangian} \end{multline}

\begin{figure}[h] \begin{center}
\includegraphics{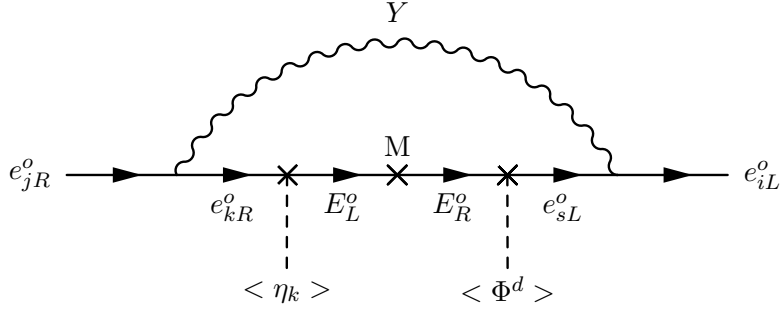}
\caption{ Generic one loop diagram contribution to the mass term
$m_{ij} \:{\bar{e}}_{iL}^o e_{jR}^o$.}
\end{center} \end{figure}

\noindent where $g_H$ is the $SU(3)$ coupling constant, $Z_1$, $Z_2$
and $Y_i^j\;,i=1,2,3\;,j=1,2$ are the eight gauge bosons. The
crosses in the internal fermion line mean tree level mixing, and
the mass $M$ generated by the Yukawa couplings in Eq.(\ref{DiracYC})
after the scalar fields get VEV's. The one loop diagram of Fig.1
gives the generic contribution to the mass term $m_{ij}
\:{\bar{e}}_{iL}^o e_{jR}^o$

\begin{equation} c_Y \frac{\alpha_H}{\pi} \sum_{k=3,4} m_k^o
\:(V_L^o)_{ik}(V_R^o)_{jk} f(M_Y, m_k^o) \qquad , \qquad \alpha_H
\equiv \frac{g_H^2}{4 \pi} \end{equation}

\noindent  where $M_Y$ is the gauge boson mass, $c_Y$ is a factor
coupling constant, Eq.(\ref{SU3lagrangian}), $m_3^o=-\sqrt{\lambda_-}$ and
$m_4^o=\sqrt{\lambda_+}$ are the See-saw mass eigenvalues,
Eq.(\ref{tleigenvalues}), and $f(x,y)=\frac{x^2}{x^2-y^2}
\ln{\frac{x^2}{y^2}}$. Using the results of Appendix A, we
compute

\begin{equation} \sum_{k=3,4} m_k^o \:(V_L^o)_{ik}(V_R^o)_{jk} f(M_Y,
m_k^o)= \frac{a_i \:b_j \:M}{\lambda_+ - \lambda_-}\:
F(M_Y) \:,\end{equation}

\noindent $i,j=1,2,3$ and $F(M_Y)\equiv
\frac{M_Y^2}{M_Y^2 - \lambda_+} \ln{\frac{M_Y^2}{\lambda_+}} -
\frac{M_Y^2}{M_Y^2 - \lambda_-} \ln{\frac{M_Y^2}{\lambda_-}}$. Adding up all
the one loop $SU(3)$ gauge boson contributions, we get the mass terms
$\bar{\psi^{o}_L} {\cal{M}}_1^o  \:\psi^{o}_R + h.c.$,

\begin{equation} {\cal{M}}_1^o = \begin{pmatrix}
R_{11} & R_{12} & R_{13}  & 0\\
0 & R_{22} & R_{23} & 0\\
0 & R_{32} & R_{33} & 0\\
0 & 0 & 0 & 0
\end{pmatrix} \;,
\end{equation}

\begin{eqnarray}
R_{11}=\frac{1}{2} {\mu}_{22} F_1 + \frac{1}{2} {\mu}_{33} F_2  &  R_{12}=( - \frac{1}{4} G_1 + \frac{1}{12} G_2 ) \:{\mu}_{12} \nonumber \end{eqnarray}

\begin{eqnarray} R_{22}=\frac{1}{4} \:{\mu}_{22} G_1 + \frac{1}{2} {\mu}_{33} F_3 + \frac{1}{12} {\mu}_{22} G_2 - G_m {\mu}_{22} \end{eqnarray}

\begin{eqnarray} R_{33}= \frac{1}{2} \:{\mu}_{22} F_3 + \frac{1}{3} \:{\mu}_{33} G_2 \quad ,\quad
R_{13}= - \frac{1}{6}\:{\mu}_{13} G_2 - {\mu}_{13} G_m \:,\nonumber            \end{eqnarray}

\begin{eqnarray} R_{23}= - \frac{1}{6}\:{\mu}_{23} G_2 + {\mu}_{23} G_m  \quad ,\quad R_{32}= - \frac{1}{6}\:{\mu}_{32} G_2 + {\mu}_{32} G_m  \;. \nonumber \end{eqnarray}

\vspace{3mm} \noindent Here,

\begin{equation*}  F_1 \equiv  \frac{\alpha_2}{\pi} F(M_1)  \quad , \quad
F_2 \equiv  \frac{\alpha_3}{\pi} F(M_2) \quad , \quad F_3 \equiv  \frac{\alpha_3}{\pi} F(M_3)
\end{equation*}

\begin{equation*}
G_1 \equiv  \frac{\alpha_2}{\pi} G_{Z_1} \qquad , \qquad  G_2 \equiv  \frac{\alpha_3}{\pi} G_{Z_2} \end{equation*}

\begin{equation*}
G_m \equiv  \frac{\sqrt{\alpha_2 \alpha_3} }{\pi} \frac{\cos\phi \sin\phi[F(M_-) - F(M_+) ] }{2\sqrt{3}}
\end{equation*}

\begin{equation*}
G_{Z_1}=\cos\phi \,F(M_-) - \sin\phi \,F(M_+) \; , \; G_{Z_2}=\sin\phi \,F(M_-) + \cos\phi \,F(M_+) \; ,
\end{equation*}

\vspace{4mm}
\noindent where $M_1 \:,\:M_2\:,\:M_3$, $M_-^2$, and  $M_+^2$ are the horizontal boson masses defined in Eqs.(\ref{M1M2},\ref{MmMp} ),

\begin{equation} {\mu}_{i j}=\frac{a_i \:b_j \:M}{\lambda_+ - \lambda_-} = \frac{a_i
\:b_j}{a \:b} \:\sqrt{\lambda_-}\:c_{\alpha} \:c_{\beta}  \quad , \quad
\alpha_2 = \frac{g_{H_2}^2}{4 \pi}  \quad , \quad \alpha_3 = \frac{g_{H_3}^2}{4 \pi}\end{equation}

\vspace{3mm}
\noindent and $c_{\alpha} \equiv \cos\alpha \:,\;c_{\beta} \equiv \cos\beta \:,\;
s_{\alpha} \equiv \sin\alpha \:,\;s_{\beta} \equiv \sin\beta$, as defined in the
Appendix, Eq.(\ref{Seesawmixing}). Therefore, up to one loop
corrections we obtain the fermion masses

\begin{equation} \bar{\psi}^{o}_L {\cal{M}}^{o} \:\psi^{o}_R + \bar{\psi^{o}_L}
{\cal{M}}_1^o \:\psi^{o}_R = \bar{\chi_L} \:{\cal{M}}
\:\chi_R \:,\end{equation}

\vspace{1mm} \noindent with ${\cal{M}} \equiv  \left[ Diag(0,0,-
\sqrt{\lambda_-},\sqrt{\lambda_+})+ {V_L^o}^T {\cal{M}}_1^o
\:V_R^o \right]$.

\pagebreak
Using $V_L^o$, $V_R^o$ from Eqs.(\ref{VoVI}) we get the mass matrix in
Version I:

\begin{equation} {\cal{M}}= \left( \begin{array}{rrcc} m_{11}&m_{12}&c_\beta \:m_{13}&s_\beta \:m_{13} \\
                                                             \\
m_{21}& m_{22} & c_\beta \:m_{23} & s_\beta \:m_{23}\\
                                                             \\
c_\alpha \:m_{31}& c_\alpha \:m_{32} & (-\sqrt{\lambda_-}+c_\alpha c_\beta
\:m_{33}) & c_\alpha s_\beta \:m_{33} \\
                                           \\
s_\alpha \:m_{31}& s_\alpha \:m_{32} & s_\alpha c_\beta \:m_{33} &
(\sqrt{\lambda_+}+s_\alpha s_\beta \:m_{33})
\end{array} \right) \;,\label{massVI}\end{equation}

\vspace{4mm} \noindent where the mass entries $m_{ij}\: ;i,j=1,2,3$ are written
as:

\begin{eqnarray}
m_{11}=\frac{1}{2} \frac{a_2}{a^\prime}  \Pi_1 \quad ,& \quad m_{12}= - \frac{1}{2} \frac{a_1 b_3}{a^\prime b} (  \Pi_2 -6 \mu_{22} G_m )  \\
                    \nonumber \\
m_{21}= \frac{1}{2} \frac{a_1 a_3}{a^\prime a}\Pi_1  \quad ,& \quad
m_{31}=\frac{1}{2} \frac{a_1}{a}  \Pi_1 \end{eqnarray}

\begin{equation} m_{13}=- \frac{1}{2} \frac{a_1 b_2}{a^\prime b} [\Pi_2 +2(2\frac{b_3^2}{b_2^2}-1)
\mu_{22}G_m ]   \end{equation}

\begin{equation}
m_{22}=\frac{1}{2} \frac{a_3 b_3}{a \, b} \left[\frac{a_2}{a^\prime} ( \Pi_2 -6 \mu_{22} G_m )+
\frac{a^\prime b_2}{a_3 b_3} (  \Pi_3 + \Delta )        \right]
\end{equation}

\begin{equation} m_{23}=\frac{1}{2} \frac{a_3 b_3}{a \, b}  \left[\frac{a_2 b_2}{a^\prime b_3} ( \Pi_2 +2(2\frac{b_3^2}{b_2^2}-1 ) \mu_{22} G_m ) - \frac{a^\prime}{a_3} (  \Pi_3 -\frac{b_2^2}{b_3^2} \Delta +2\frac{b^2}{b_3^2}\mu_{33} G_m    )        \right] \end{equation}

\begin{equation} m_{32}=\frac{1}{2} \frac{a_3 b_3}{a \, b}  \left[\frac{a_2}{a_3} ( \Pi_2 -6 \mu_{22} G_m)-\frac{b_2}{b_3} (  \Pi_3 -\frac{{a^\prime}^2 }{a_3^2} \Delta -2\frac{a^2}{a_3^2}\mu_{33} G_m ) \right] \end{equation}

\begin{equation}
m_{33}=\frac{1}{2} \frac{a_3 b_3}{a \, b} \left[\frac{a_2 b_2}{a_3 b_3} ( \Pi_2 - 2 \mu_{22} G_m ) + \Pi_3+ \frac{ {a^\prime}^2 b_2^2}{a_3^2 b_3^2} \Delta - \frac{1}{3} \frac{a^2 b^2}{a_3^2 b_3^2}\mu_{33} G_2  + 2 ( \frac{b_2^2}{b_3^2} + 2\frac{a_2^2}{a_3^2}-\frac{{a^\prime}^2}{a_3^2} )\mu_{33} G_m          \right]
\end{equation}

\vspace{3mm}
\begin{eqnarray}
\Pi_1 = \mu_{22} F_1 +  \mu_{33} F_2 \quad ,& \quad \Pi_2 = \mu_{22} G_1 +  \mu_{33} F_3 \nonumber \\
                                                                             \nonumber \\
\Pi_3 = \mu_{22} F_3 +  \mu_{33} G_2 \quad ,& \quad \Delta = \frac{1}{2}\mu_{33}(G_2 - G_1 )
\end{eqnarray}

\begin{equation*} a^\prime=\sqrt{a_1^2+a_2^2} \;\;, \;\;a=\sqrt{{a^\prime}^2+a_3^2} \;\; ,
\;\; b=\sqrt{b_2^2+b_3^2} \;, \end{equation*}

\vspace{4mm}
Notice that the $m_{ij}$ mass terms up to one loop depend just on the ratios
$\frac{a_i}{a_j}=\frac{v_i}{v_j} \,( \frac{V_i}{V_j} ) \,$ and $\frac{b_2}{b_3}=\frac{h_2 \,\Lambda_2}{h_3 \,\Lambda_3}$ coming from the tree level see-saw mass matrix $\mathcal{M}^o$.

\vspace{5mm}

\noindent For $V_L^o$, $V_R^o$ from Eqs.(\ref{VoVII}) we get the Version II:

\begin{equation} {\cal M}= \left( \begin{array}{rrcc} M_{11}&M_{12}&c_\beta \:M_{13}&s_\beta \:M_{13} \\
                                                             \\
M_{21}& M_{22} & c_\beta \:M_{23} & s_\beta \:M_{23}\\
                                                             \\
c_\alpha \:M_{31}& c_\alpha \:M_{32} & (-\sqrt{\lambda_-}+c_\alpha c_\beta
\:M_{33}) & c_\alpha s_\beta \:M_{33} \\
                                           \\
s_\alpha \:M_{31}& s_\alpha \:M_{32} & s_\alpha c_\beta \:M_{33} &
(\sqrt{\lambda_+}+s_\alpha s_\beta \:M_{33})
\end{array} \right) \;,\label{massVII} \end{equation}

\vspace{4mm} \noindent where the mass terms $M_{ij}\: ;i,j=1,2,3$ may be obtained from those of
 $m_{ij}$ as follows

\begin{equation} \begin{array}{ccc}
M_{11}=m_{22}  ,& M_{12}=m_{21} , & M_{13}=m_{23} \\
                                                   \\
M_{21}=-m_{12} ,& M_{22}=- m_{11} , & M_{23}=- m_{13} \\
                                                   \\
M_{31}=m_{32} ,& M_{32}=m_{31} , & M_{33}=m_{33} \end{array}
\end{equation}

\vspace{4mm}

\noindent The diagonalization of ${\cal{M}}$,
Eq.(\ref{massVI}) or Eq.(\ref{massVII}), gives the physical masses for  u,
d, e charged fermions. Neutrinos may also get left and right handed Majorana masses.

Using a new biunitary transformation
$\chi_{L,R}=V_{L,R}^{(1)} \;\Psi_{L,R}$;
\;$\bar{\chi}_L \;{\cal{M}} \;\chi_R= \bar{\Psi}_L \:{V_L^{(1)}}^T
{\cal{M}} \; V_R^{(1)} \:\Psi_R $, with ${\Psi_{L,R}}^T = ( f_1 ,
f_2 , f_3 , F )_{L,R}$ the mass eigenfields, that is

\begin{equation}
{V^{(1)}_L}^T {\cal{M}} \:{\cal M}^T \;V^{(1)}_L =
{V^{(1)}_R}^T {\cal M}^T \:{\cal{M}} \;V^{(1)}_R =
Diag(m_1^2,m_2^2,m_3^2,M_F^2) \:,\end{equation}

\noindent $m_1^2=m_e^2$, $m_2^2=m_\mu^2$, $m_3^2=m_\tau^2$ and
$M_F^2=M_E^2$ for charged leptons. Therefore, the transformation from
massless to massive fermion eigenfields read

\begin{equation} \psi_L^o = V_L^{o} \:V^{(1)}_L \:\Psi_L \qquad \mbox{and}
\qquad \psi_R^o = V_R^{o} \:V^{(1)}_R \:\Psi_R \end{equation}

\subsection{Quark Mixing and non-unitary $( V_{CKM} )_{4\times 4}$ }

Recall that vector like quarks, Eq.(\ref{vectorquarks}), are $SU(2)_L$
weak singlets, and then they do not couple to $W$ boson in the
interaction basis. So, the interaction of ordinary quarks ${f_{uL}^o}^T=(u^o,c^o,t^o)_L$ and
${f_{dL}^o}^T=(d^o,s^o,b^o)_L$ to the $W$ charged gauge boson
is

\begin{equation} \frac{g}{\sqrt{2}} \,\bar{f^o}_{u L} \gamma_\mu f_{d L}^o
{W^+}^\mu = \frac{g}{\sqrt{2}} \,\bar{\Psi}_{u L}\;{V_{u L}^{(1)}}^T\;[(V_{u
L}^o)_{3\times 4}]^T \;(V_{d L}^o)_{3\times 4} \;V_{d
L}^{(1)}\;\gamma_\mu \Psi_{d L} \;{W^+}^\mu \:,\end{equation}

\noindent with $g$ is the $SU(2)_L$ gauge coupling. Hence, the non-unitary $V_{CKM}$ of dimension $4\times
4$ is identified as

\begin{equation} (V_{CKM})_{4\times 4}\equiv {V_{u L}^{(1)}}^T\;[(V_{u
L}^o)_{3\times 4}]^T \;(V_{d L}^o)_{3\times 4} \;V_{d L}^{(1)} \equiv {V_{u L}^{(1)}}^T\; V_o \;V_{d L}^{(1)}
\:.\end{equation}

\noindent For instance, for u and d quark masses $\mathcal{M}$ in Eq.(\ref{massVII});

\small
\begin{equation} V_o=\begin{pmatrix}
\frac{v_3 V_3}{v V} co + \frac{v^\prime V^\prime}{v V} &-\frac{v_3}{v} so  & c_\alpha^d (\frac{v_3 V^\prime}{v V} co - \frac{v^\prime V_3}{v V} ) & s_\alpha^d (\frac{v_3 V^\prime}{v V} co - \frac{v^\prime V_3}{v V} )  \\
 &   &   &   \\
\frac{V_3}{V} so & co  & \frac{V^\prime}{V} c_\alpha^d so       &   \frac{V^\prime}{V} s_\alpha^d so                                              \\
&   &   &   \\
c_\alpha^u (\frac{v^\prime V_3}{v V} co - \frac{v_3 V^\prime}{v V} ) & - \frac{v^\prime}{v} c_\alpha^u so  & c_\alpha^d c_\alpha^u (\frac{v^\prime V^\prime}{v V} co + \frac{v_3 V_3}{v V}) & c_\alpha^u s_\alpha^d (\frac{v^\prime V^\prime}{v V} co + \frac{v_3 V_3}{v V})         \\
&   &   &   \\
s_\alpha^u (\frac{v^\prime V_3}{v V} co - \frac{v_3 V^\prime}{v V} ) & - \frac{v^\prime}{v} s_\alpha^u so & c_\alpha^d s_\alpha^u (\frac{v^\prime V^\prime}{v V} co + \frac{v_3 V_3}{v V}) & s_\alpha^u s_\alpha^d (\frac{v^\prime V^\prime}{v V} co + \frac{v_3 V_3}{v V})
\end{pmatrix}
\end{equation}

\normalsize
\begin{equation} s_o=\frac{v_1}{v^\prime}\:\frac{V_2}{V^\prime} - \frac{v_2}{v^\prime}\:\frac{V_1}{V^\prime}
\quad , \quad
c_o=\frac{v_1}{v^\prime}\:\frac{V_1}{V^\prime} +
\frac{v_2}{v^\prime}\:\frac{V_2}{V^\prime} \label{somixing} \end{equation}

\begin{equation} c_o^2+s_o^2=1       \end{equation}

\section{Preliminary numerical results for charged leptons} \label{numericale}

\subsection{Charged leptons:}

Using the strong hierarchy for quarks and charged leptons masses
and the results in\cite{prd2007}, we report here the magnitudes of
lepton masses and mixing coming from a parameter space fit in this model.

In the approach $\Lambda_2 \approx \Lambda_3 = \Lambda$ or $\frac{M_2}{M_1} \approx \mathcal{O}(1)$:

\begin{equation*} M_1 \approx  M_2 = M \quad , \quad  M^2=\frac{g_H^2 \Lambda^2}{2} \quad ,
\quad M_-^2 \approx \frac{2}{3} M^2 \quad , \quad M_-^2 \approx 2 M^2 \end{equation*}

\noindent Using the parameters from the $SU(3)$ flavour symmetry
\begin{equation*} \frac{\alpha_H}{\pi}=0.06 \quad , \quad M=10\,\text{TeV}  \end{equation*}

\noindent and the ones coming from the tree level seesaw

\begin{equation*} \sin\alpha=0.1 \; , \; \sin\beta=0.0081 \; , \; \sqrt{\lambda_-}=2138.22 \,\text{MeV} \; ,\; \sqrt{\lambda_+}=2.6 \,\text{TeV} \end{equation*}

\begin{equation*} \frac{V_1}{V_2} \simeq 0.1 \quad , \quad \frac{V_2}{\sqrt{V_1^2+V_2^2}} \simeq 0.995037
\end{equation*}

\begin{equation*}
\frac{\sqrt{V_1^2+V_2^2}}{V_3} \simeq 0.344266  \quad , \quad \frac{h_3}{h_2} \simeq -0.218152
\end{equation*}

\noindent the mass matrix

\begin{equation} M_e=
\begin{pmatrix}
1.0558  & -9.6693  & -62.5208  & -0.5115  \\
                                        \\
-1.6717 & 102.263  & 7.7087  & 0.0630   \\
                                         \\
39.19 & -3.3121  & -1744.59  & 3.2209  \\
                                        \\
3.9387 & -0.3328 & 39.5616  & 2.6 \times 10^6
\end{pmatrix} \end{equation}

\noindent fit the charged lepton masses at the $M_Z$ scale \cite{xingzhang}:

\begin{equation*}
\left( m_e \,,\, m_\mu \,,\, m_\tau \, \right) = \left( 0.486 \,,\,102.7\,,\,1746.17\,\right)\,\text{MeV}
\end{equation*}

\noindent and a mass for the vector like electron of $M_E \approx 2.6\,\text{TeV}$ with mixing matrix

\begin{equation} V_{e \,L}^{(1)}=
\begin{pmatrix}
0.9950 & -0.0926  & -0.0358  & -1.9713 \times 10^{-7} \\
                                                     \\
0.0928 & 0.9956  & 0.0045  & 2.4300 \times 10^{-8}    \\
                                                     \\
-0.0352 &  0.0078 & -0.9993  & 1.2286 \times 10^{-6}  \\
                                                      \\
 2.3719 \times 10^{-7}& -5.2139 \times 10^{-8} & 1.2206 \times 10^{-6}   & 1
\end{pmatrix} \end{equation}

\vspace{7mm}
\section{Tree level neutrino masses}

\vspace{5mm}
\subsection{Tree level Dirac Neutrino masses}

Here we present a preliminary study of neutrino masses. For neutrinos
we may write the Dirac type gauge invariant couplings

\begin{equation} h_D \,\bar{\Psi}_l^o \,\Phi^u \,N_R^o \;\;+\;\; h_2 \,\bar{\Psi}_\nu^o
\,\eta_2 \,N_L^o \;\;+\;\; h_3 \,\bar{\Psi}_\nu^o \,\eta_3 \,N_L^o \;\;+\;\ M_D \,\bar{N}_L^o \,N_R^o \;\;+ h.c
\end{equation}

\noindent $h$, $h_2$ and $h_3$ are Yukawa couplings, and $M_D$  a Dirac type
invariant neutrino mass for the sterile neutrino $N_{L,R}^O$. After electroweak
symmetry breaking, we obtain in the interaction basis $\Psi_{L,R}^{o\, T} = ( \nu_e , \nu_\mu , \nu_\tau , N )_{L,R}^o$,  the mass terms

\begin{equation} \overline{\Psi_{\nu}^o}_L  \;{\mathcal{M}}_{\nu \,D}^o \;\Psi_{\nu \,R}^o
\;\;+ h.c \, ,\nonumber \end{equation}

\noindent where

\begin{equation} {\mathcal{M}}_D^{\nu \,(o)} =\begin{pmatrix}
0 & 0 & 0 & h_D \:v_1\\
0 & 0 & 0 & h_D \:v_2\\
0 & 0 & 0 & h_D \:v_3\\
0 &  h_2 \Lambda_2 & h_3 \Lambda_3 & M_D
\end{pmatrix}  \end{equation}

\vspace{5mm}
\subsection{Tree level Majorana masses:}

 Since $N_{L,R}^o$, Eq.(\ref{vectorleptons}), are completely sterile neutrinos, we may also write the left and right handed Majorana type couplings

\begin{equation} h_L \,\overline{\Psi_l^o} \,\Phi (N_L^o)^c  \quad + \quad m_L \,\overline{N_L^o}\, (N_L^o)^c \end{equation}

\noindent and

\begin{equation} h_{2 R} \,\overline{(\Psi_\nu^o)^c} \,\eta_3^\dag \,N_R^o \;\;+\;\;h_{3 R} \,\overline{(\Psi_\nu^o)^c}
\,\eta_2^\dag \,N_R^o  \;\;+\;\;m_R \,\overline{N_R^o} \,(N_R^o)^c + h.c \; ,\end{equation}

\noindent respectively. After spontaneous symmetry breaking, we also get the left handed and right handed Majorana mass terms

\begin{eqnarray} h_L \,\left[  v_ 1 \,\overline{\nu_{e L}^o} + v_ 2 \,\overline{\nu_{\mu L}^o} + v_ 3 \, \overline{\nu_{\tau L}^o}         \right] \,(N_L^o)^c  \quad +
\quad m_L \,\overline{N_L^o} \,(N_L^o)^c   \nonumber     \\
                                                               \nonumber  \\
+ \left[  h_{2 R} \,\Lambda_2 \,\overline{(\nu_{\mu R}^o)^c} + h_{3 R} \,\Lambda_3 \,\overline{(\nu_{\tau R}^o)^c }  \right] \,N_R^o   +  \;\;+\;\;m_R \,\overline{N_R^o} \,(N_R^o)^c + h.c. \, ,
\end{eqnarray}

\noindent and the Generic Majorana mass matrix for neutrinos may be written as

\begin{equation}
\begin{pmatrix} \overline{\Psi_\nu^o}_L \;,& \overline{\Psi_\nu^{o c}}_L \end{pmatrix}
\,\mathcal{M}_\nu \,\begin{pmatrix} \Psi_{\nu \,R}^{o \,c}  \\
                                                         \\
\Psi_{\nu \,R}^o \end{pmatrix}  \; , \; \Psi_{\nu \,L,R}^o = \begin{pmatrix} \nu_i^o   \\
N^o \end{pmatrix}_{L,R} \; , \;
\Psi_{\nu \,L,R}^{o \,c}  = \begin{pmatrix} \nu_i^{o \,c}   \\
N^{o \,c} \end{pmatrix}_{L,R}
\end{equation}

\noindent where

\begin{equation} \mathcal{M}_\nu^{(o)} = \frac{1}{2} \begin{pmatrix} M_{\nu \,L}^o & M_{\nu \,D}^o  \\
                                               \\
( M_{\nu \,D}^o )^T   & M_{\nu \,R}^o
\end{pmatrix}
\label{nuMajorana} \end{equation}

\noindent and
\begin{equation} M_{\nu \,L}^o = \begin{pmatrix}
 0 & 0 & 0 & h_L v_1   \\
                        \\
 0 & 0 & 0 & h_L v_2 \\
                        \\
 0 & 0 & 0 & h_L v_3  \\
                        \\
 h_L v_1  & h_L v_2 & h_L v_3 & m_L
\end{pmatrix}  \; , \; M_{\nu \,R}^o= \begin{pmatrix}
 0 & 0 & 0 & 0  \\
                \\
 0 & 0 & 0 & h_{2 R} \Lambda_2  \\
                                    \\
 0 & 0 & 0 & h_{3 R} \Lambda_3  \\
                                    \\
 0 & h_{2 R} \Lambda_2  & h_{3 R} \Lambda_3  & m_R
\end{pmatrix} \;, \end{equation}

\vspace{5mm}
\section{One loop neutrino masses}

Diagonalization of the $8\times 8$  Majorana mass matrix $\mathcal{M}_\nu^{(o)}$ in
Eq.(\ref{nuMajorana}) yields four zero eigenvalues at tree level. Three of these
massless neutrinos correspond to the active ordinary neutrinos. So, in this scenario
light neutrinos may get very small masses from radiative corrections mediated
by the $SU(3)$ heavy gauge bosons.

\vspace{5mm}
\subsection{One loop Dirac Neutrino masses}

Neutrinos may get tiny Dirac mass terms from the generic
one loop diagram in Fig. 2, as well as L-handed and R-handed Majorana masses from
Fig. 3 and Fig. 4, respectively. The contribution from these diagrams read

\begin{equation} c_Y \frac{\alpha_H}{\pi} \sum_{k=1,2,3,4} m_k^o
\:U^o_{ik} U^o_{jk} f(M_Y, m_k^o) =  c_Y \frac{\alpha_H}{\pi} \,m_\nu(M_Y)_{ij}   \quad , \quad \alpha_H
\equiv \frac{g_H^2}{4 \pi} \end{equation}

\noindent with
\begin{equation}m_\nu(M_Y)_{ij} \equiv \sum_{k=1,2,3,4} m_k^o \:U^o_{ik} U^o_{jk}\:f(M_Y, m_k^o) ,\end{equation}

\begin{figure}[h] \begin{center}
\includegraphics{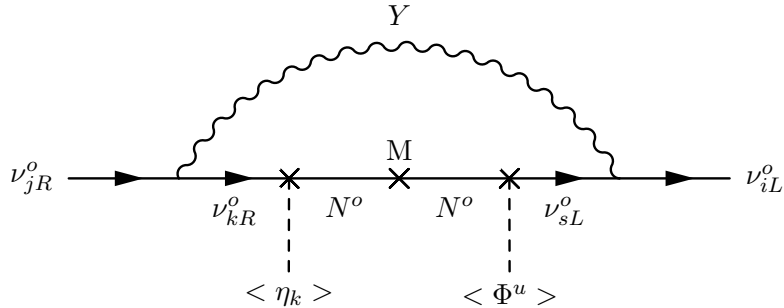}
\caption{ Generic one loop diagram contribution to the Dirac mass term
$m_{ij} \:{\bar{\nu}}_{iL}^o \nu_{jR}^o$. $\; \text{M}=M_D , m_L, m_R$}
\end{center} \end{figure}

\begin{equation}  {\mathcal{M}}_{\nu \,D}^{(1)} = \begin{pmatrix}
R_{\nu \,15}  & R_{\nu \,16}  & R_{\nu \,17}   & 0 \\
                                                                        \\
0 & R_{\nu \,26}  & R_{\nu \,27}  & 0 \\
                                                                          \\
0  &  R_{\nu \,36}  & R_{\nu \,37} & 0    \\
                                                                           \\
0 & 0 & 0 & 0
\end{pmatrix} \end{equation}

\begin{equation*}
R_{\nu \,15} =\frac{1}{2} m_\nu(M_1)_{26} + \frac{1}{2} m_\nu(M_2)_{37}  \; , \;
R_{\nu \,16} = - \frac{1}{4} m_\nu(M_{Z_1})_{16} + \frac{1}{12} m_\nu(M_{Z_2})_{16} \;,
\end{equation*}

\begin{equation*} R_{\nu \,26} =\frac{1}{4} \:m_\nu(M_{Z_1})_{26} + \frac{1}{2} m_\nu(M_3)_{37} + \frac{1}{12} m_\nu(M_{Z_2})_{26} - G_{\nu,m \,26} \;,\end{equation*}

\begin{equation*} R_{\nu \,37} = \frac{1}{2} m_\nu(M_3)_{26}\: + \frac{1}{3} m_\nu(M_{Z_2})_{37} \; ,\;
R_{\nu \,17} = - \frac{1}{6} m_\nu(M_{Z_2})_{17} - G_{\nu,m \,17}
 \:,\end{equation*}

\begin{equation*} R_{\nu \,27} = - \frac{1}{6}\:m_\nu(M_{Z_2})_{27} + G_{\nu,m \,27}
  \; ,\;
R_{\nu \,36} = - \frac{1}{6}\:m_\nu(M_{Z_2})_{36} + G_{\nu,m \,36}  \;,  \end{equation*}

\begin{eqnarray}
m_\nu(M_{Z_1})_{ij}= \cos\phi \,m_\nu(M_-)_{ij} - \sin\phi \,m_\nu(M_+)_{ij}     \nonumber \\
                                                     \nonumber \\
m_\nu(M_{Z_2})_{ij}= \sin\phi \,m_\nu(M_-)_{ij} + \cos\phi \,m_\nu(M_+)_{ij}       \nonumber \\
                                                      \nonumber \\
G_{\nu,m \,ij}=  \frac{\sqrt{\alpha_2 \alpha_3} }{\pi} \,\frac{1}{2\sqrt{3}}
\,\cos\phi \,\sin\phi \,[m_\nu(M_-)_{ij} - m_\nu(M_+)_{ij}  ]
\end{eqnarray}

\vspace{5mm}
\subsection{One loop L-handed Majorana masses}

\begin{figure}[h] \begin{center}
\includegraphics{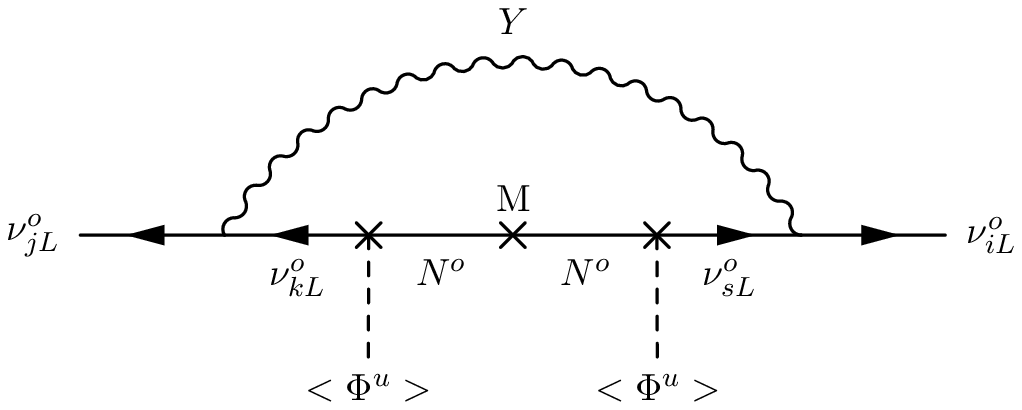}
\caption{ Generic one loop diagram contribution to the L-handed Majorana mass term
$m_{ij} \:{\bar{\nu}}_{iL}^o (\nu_{jL}^o)^T$. $\; \text{M}=M_D , m_L, m_R$}
\end{center} \end{figure}

\begin{equation}  M_{\nu \,L}^{(1)} = \begin{pmatrix}
 m_{\nu \,11} & m_{\nu \,12} &  m_{\nu \,13} & 0  \\
                                                                        \\
 m_{\nu \,12} & m_{\nu \,22}  & m_{\nu \,23}  & 0  \\
                                                                          \\
 m_{\nu \,13}  & m_{\nu \,23}  & m_{\nu \,33}  & 0   \\
                                                                           \\
0 & 0 & 0 & 0
\end{pmatrix} \end{equation}

\begin{eqnarray}
m_{\nu \,11}=\frac{1}{4} \,m-\nu(M_{Z_1})_{11} + \frac{1}{12} \,m_\nu(M_{Z_2})_{11} +
G_{\nu,m \,11}               \nonumber \\
                               \nonumber \\
m_{\nu \,12}=-\frac{1}{4} \,m_\nu(M_{Z_1})_{12} + \frac{1}{12} \,m_\nu(M_{Z_2})_{12} +
\frac{1}{2} \,m_\nu(M_1)_{12}  \nonumber \\
                               \nonumber \\
m_{\nu \,13}=\frac{1}{2} \,m-\nu(M_2)_{13} - \frac{1}{6} \,m_\nu(M_{Z_2})_{13} -
G_{\nu,m \,13}              \nonumber \\
                               \nonumber \\
m_{\nu \,22}=\frac{1}{4} \,m_\nu(M_{Z_1})_{22} + \frac{1}{12} \,m_\nu(M_{Z_2})_{22} -
G_{\nu,m \,22}               \nonumber \\
                               \nonumber \\
m_{\nu \,23}=\frac{1}{2} \,m_\nu(M_3)_{23} - \frac{1}{6} \,m_\nu(M_{Z_2})_{23} +
G_{\nu,m \,23}               \nonumber \\
                               \nonumber \\
m_{\nu \,33}= \frac{1}{3} \,m_\nu(M_{Z_2})_{33}
\end{eqnarray}

\vspace{5mm}
\subsection{One loop R-handed Majorana masses}

\begin{figure}[h] \begin{center}
\includegraphics{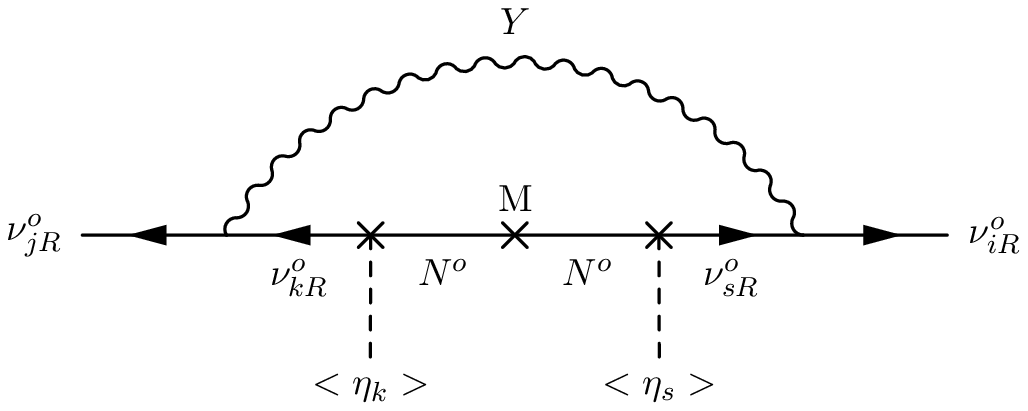}
\caption{ Generic one loop diagram contribution to the R-handed Majorana mass term
$m_{ij} \:{\bar{\nu}}_{iR}^o (\nu_{jR}^o)^T$. $\; \text{M}=M_D , m_L, m_R$}
\end{center} \end{figure}

\begin{equation}  M_{\nu \,R}^{(1)} = \begin{pmatrix}
 0 & 0 &  0 & 0  \\
                                                                        \\
 0 & m_{\nu \,66}  & m_{\nu \,67}  & 0  \\
                                                                          \\
 0 & m_{\nu \,67}  & m_{\nu \,77}  & 0   \\
                                                                           \\
0 & 0 & 0 & 0
\end{pmatrix} \end{equation}

\begin{eqnarray}
m_{\nu \,66}=\frac{1}{4} \,m_\nu(M_{Z_1})_{66} + \frac{1}{12} \,m_\nu(M_{Z_2})_{66} -
G_{\nu,m \,66}               \nonumber \\
                               \nonumber \\
m_{\nu \,67}=\frac{1}{2} \,m_\nu(M_3)_{67} - \frac{1}{6} \,m_\nu(M_{Z_2})_{67} +
G_{\nu,m \,67}               \nonumber \\
                               \nonumber \\
m_{\nu \,77}= \frac{1}{3} \,m_\nu(M_{Z_2})_{77}
\end{eqnarray}

\section{Conclusions}

We have reported recent analysis on charged lepton and neutrino masses
within a $SU(3)$ family symmetry model extension, which combines tree level "Dirac
See-saw" mechanisms and radiative corrections to implement a
successful hierarchical spectrum for charged fermion masses and quark mixing.

We report in section \ref{numericale} a preliminary solution for ordinary charged lepton masses
$( m_e \,,\, m_\mu \,,\, m_\tau ) = ( 0.486 \,,\,102.7\,,\,1746.17 )\,\text{MeV}$ at the $M_Z$
scale with a vector-like electron mass $M_E \approx 2.6 \,TeV$ and horizontal gauge boson masses
of the order of 10 TeV. A numerical fit on neutrino masses and mixing is in progress and results
will be reported elsewhere.

\vspace{3mm}
It is worth to mention that due to the assigned of the ordinary fermions
as triplets under $SU(3)$, each type of u, d, e and neutrinos couples
just to one of the $\Phi^u$ or $\Phi^d$ scalars. So, u-quarks  and neutrinos
coupled to $\Phi^u$ while d-quarks and charged leptons couple to $\Phi^d$.
The scalar fields introduced to break the symmetries in the model:
$\Phi$, $\Phi^\prime$, $\eta_1$, $\eta_2$ and $\eta_3$, couple to ordinary fermions
through the Eq.(\ref{DiracYC}). Therefore, FCNC scalar couplings to ordinary fermions are suppressed
by light-heavy mixing angles, which may be small enough to suppress properly the FCNC
mediated by the scalar fields within this scenario.

\vspace{5mm}
\section*{Acknowledgements}

I wish to thank all the organizers and colleagues, in particular
to N.S. Mankoc-Borstnik, M. Y. Khlopov and H.B. Nielsen for very useful discussions, and for the
stimulating Workshop at Bled, Slovenia. This work was
partially supported by the "Instituto Polit\'ecnico Nacional",
(Grants from EDI and COFAA) and "Sistema Nacional de
Investigadores" (SNI) in Mexico.


\appendix

\section{Diagonalization of the generic Dirac See-saw mass matrix}

\begin{equation} {\cal M}^o=
\begin{pmatrix} 0 & 0 & 0 & a_1\\ 0 & 0 & 0 & a_2\\ 0 & 0 & 0 &
a_3\\ 0 & b_2 & b_3 & c \end{pmatrix} \end{equation}

\vspace{1mm} \noindent Using a biunitary transformation
$\psi^{o}_L = V_L^o \:\chi_L$ and  $\psi^{o}_R = V_R^o
\:\chi_R $ to diagonalize ${\cal{M}}^o$, the orthogonal matrices
$V^{o}_L$ and $V^{o}_R$ may be written explicitly as the following
two versions

\vspace{4mm}
Version I:
\vspace{3mm}

\begin{equation} V^{o}_L = \begin{pmatrix} \frac{a_2}{a^\prime}& \frac{a_1 a_3}{a^\prime
a} & \frac{a_1}{a} \cos\alpha &
\frac{a_1}{a} \sin\alpha\\
                        \\
- \frac{a_1}{a^\prime}  & \frac{a_2 a_3}{a^\prime a}  &
\frac{a_2}{a} \cos\alpha &
\frac{a_2}{a} \sin\alpha\\
                        \\
0 & - \frac{a^\prime}{a}   & \frac{a_3}{a} \cos{\alpha}
& \frac{a_3}{a} \sin{\alpha}\\
                            \\
0 & 0 & -\sin{\alpha} & \cos{\alpha}
\end{pmatrix}  \; , \; V^{o}_R = \begin{pmatrix}
1 & 0 & 0 & 0 \\
                         \\
0 & \frac{b_3}{b} & \frac{b_2}{b} \cos{\beta} & \frac{b_2}{b} \sin{\beta}\\
                                                     \\
0& - \frac{b_2}{b} & \frac{b_3}{b} \cos{\beta} & \frac{b_3}{b} \sin{\beta}\\
                         \\
0 & 0 & -\sin{\beta} & \cos{\beta}
\end{pmatrix}   \label{VoVI} \end{equation}

\vspace{5mm}

Version II:
\vspace{3mm}

\begin{equation} V^{o}_L = \begin{pmatrix} \frac{a_1 a_3}{a^\prime
a}  & - \frac{a_2}{a^\prime}    & \frac{a_1}{a} \cos\alpha &
\frac{a_1}{a} \sin\alpha\\
                        \\
\frac{a_2 a_3}{a^\prime a}  & \frac{a_1}{a^\prime}    &
\frac{a_2}{a} \cos\alpha &
\frac{a_2}{a} \sin\alpha\\
                        \\
- \frac{a^\prime}{a} &   0  & \frac{a_3}{a} \cos{\alpha}
& \frac{a_3}{a} \sin{\alpha}\\
                            \\
0 & 0 & -\sin{\alpha} & \cos{\alpha}
\end{pmatrix} \; , \; V^{o}_R = \begin{pmatrix}
0 & 1 & 0 & 0 \\
                         \\
\frac{b_3}{b} & 0 & \frac{b_2}{b} \cos{\beta} & \frac{b_2}{b} \sin{\beta}\\
                                                     \\
 - \frac{b_2}{b}& 0 & \frac{b_3}{b} \cos{\beta} & \frac{b_3}{b} \sin{\beta}\\
                         \\
0 & 0 & -\sin{\beta} & \cos{\beta}
\end{pmatrix} \;, \label{VoVII} \end{equation}

\vspace{3mm}
\begin{equation} \lambda_{\pm } = \frac{1}{2} \left( B \pm \sqrt{B^2 -4D} \right) \label{nonzerotleigenvalues}
\end{equation}

\vspace{3mm}
\noindent are the nonzero eigenvalues of
${\cal{M}}^{o} {{\cal{M}}^{o}}^T$ (${{\cal{M}}^{o}}^T
{\cal{M}}^{o}$), and

\begin{eqnarray} B = a^2 + b^2 + c^2 =
\lambda_{-}+\lambda_{+}\quad &, \quad D= a^2
b^2=\lambda_{-}\lambda_{+} \;,\label{paramtleigenvalues} \end{eqnarray}

\vspace{1mm}

 \begin{eqnarray} \cos{\alpha} =\sqrt{\frac{\lambda_+ -
a^2}{\lambda_+ - \lambda_-}} \qquad , \qquad \sin{\alpha} =
\sqrt{\frac{a^2
- \lambda_-}{\lambda_+ - \lambda_-}} \:,\nonumber \\
                              \label{Seesawmixing}  \\
\cos{\beta} =\sqrt{\frac{\lambda_+ - b^2}{\lambda_+ - \lambda_-}}
\qquad , \qquad \sin{\beta} = \sqrt{\frac{b^2 -
\lambda_-}{\lambda_+ - \lambda_-}} \:.\nonumber \end{eqnarray}

\end{document}